\documentclass[aps,prl,reprint,showpacs,superscriptaddress]{revtex4-1}
\usepackage{graphicx} 

\begin{document}
\title{Nonlinear Time-Reversal in a Wave Chaotic System }
\author{Matthew Frazier}
\affiliation{Department of Physics, University of Maryland, College Park, MD  20742-4111}
\author{Biniyam Taddese}
\author{Thomas Antonsen}
\author{Steven M. Anlage}
\affiliation{Department of Physics, University of Maryland, College Park, MD  20742-4111}
\affiliation{Department of Electrical and Computer Engineering, University of Maryland, College Park, MD  20742-3285}
\date{\today}

\begin{abstract}
Exploiting the time-reversal invariance and reciprocal properties of the lossless wave equation enables elegantly simple solutions to complex wave-scattering problems, and is embodied in the time-reversal mirror. Here we demonstrate the implementation of an electromagnetic time-reversal mirror in a wave chaotic system containing a discrete nonlinearity. We demonstrate that the time-reversed nonlinear excitations reconstruct exclusively upon the source of the nonlinearity. As an example of its utility, we demonstrate a new form of secure communication, and point out other applications. 
\end{abstract}


\maketitle

Wave chaos concerns the study of solutions to linear wave equations that display classical chaos in their short-wavelength limit.  Such systems are endowed with many universal wave properties, such as eigenvalue and scattering-matrix statistics, by virtue of their classically chaotic counterparts. \cite{a1} Although wave chaotic systems are strongly scattering and have complex behavior, they can be elegantly studied by exploiting the time-reversal invariance and reciprocal properties of the linear wave equation. \cite{a2,a3,a4,a5,a6,a7,a8,a9} Adding objects with complex nonlinear dynamics to linear wave chaotic systems has only recently been considered, \cite{a10} and represents an exciting new direction of research. Here we examine a wave chaotic system with a single discrete nonlinear element, and create a new nonlinear electromagnetic time-reversal mirror that shows promise for both fundamental studies and novel applications.

A time-reversal mirror works by taking advantage of the invariance of the lossless wave equation under time-reversal; for a time-forward solution of the wave equation representing a wave travelling in a given direction,  there is a corresponding time-reversed solution representing a wave travelling in the same direction backwards in time, or in the opposite direction forward in time. This can be realized by transmitting a waveform at a particular source location and recording the reverberating waveforms (sona) with an array of receivers; the recorded waveforms are reversed in time and retransmitted back from the receivers, propagating to and reconstructing a time-reversed version of the original waveform at the source \cite{a3}. Time-reversal mirrors have been demonstrated for both acoustic \cite{a2,a3,a4,a5,a6,a7,a8,a9, a11, a12} and electromagnetic waves \cite{a6, a8, a13}, and exploited for applications such as lithotripsy \cite{a2, a4}, underwater communication \cite{a2, a14, a15}, sensing perturbations \cite{a11, a12}, and achieving sub-wavelength imaging \cite{a6,a7,a8, a16}.  

An ideal time-reversal mirror in an open environment would collect the forward-propagating wave at every point on a closed surface enclosing the transmitter, requiring a very large number of receivers.  The receiving array can be simplified, without significant loss of fidelity of the reconstruction, if there is a closed, ray-chaotic environment where a propagating wave (with wavelength much smaller than the size of the enclosure) will eventually reach every point in the environment, allowing the use of a single receiver to capture the signal to be time-reversed \cite{a9,a11}. Reconstruction is possible even when only a small fraction of the transmitted energy is collected by the receiver.

Now consider a discrete nonlinear element added to the otherwise linear environment. When a waveform is incident on the nonlinear element, excitations are formed at frequencies different from those in the initial pulse. These new excitations appear as a new transmission originating from the nonlinear element, which in principle should be time-reversible in their propagation, and behave similarly to the initial pulse. In particular, the nonlinear excitations should, upon time-reversal, reconstruct as a well-focused signal upon the nonlinear element. Furthermore, the generation and time-reversal of nonlinear excitations will not depend upon the location of the object, which may be unknown.  This allows creation of an exclusive communication channel with the nonlinear object, without knowledge of its location, and the ability to direct energy onto it without interfering with nearby objects. 

This method of nonlinear time-reversal has been demonstrated through time-reversal of acoustic waves \cite{a17} in materials with discrete nonlinear defects, and exploited as a means of non-destructive evaluation. \cite{a18}  Nonlinear time-reversal has also been shown to be feasible in ultrasound applications, using cavitation bubbles to generate harmonics and to guide the time-reversal.\cite{b1} Time-reversible propagation of acoustic waves in a distributed nonlinear medium is also possible; \cite{a19} however, time-reversal invariance is broken for long propagation lengths in which shock waves form. For localized nonlinearities, the generated waves propagate linearly, and remain time-reversible. Time-reversal using localized nonlinearities was also demonstrated in optical systems (using phase conjugation) by two methods. In one, a nanoparticle was used as a localized nonlinearity, generating second harmonic radiation which was phase- conjugated back onto the nanoparticle. \cite{a20} In the second, a focused ultrasonic signal was used as a synthetic 'guide star' (similar to atmospheric guide stars for optical correction in astronomical imaging) for the focusing of the time-reversed light to the chosen focal point. \cite{a21}

We have realized a time-reversal mirror using electromagnetic waves at telecommunication frequencies in a closed complex (ray-chaotic) scattering environment, as shown in Fig. 1a.  The enclosure is a 1.06 m3 aluminum box with irregular surfaces and a conducting scattering paddle, and has three ports for the introduction and extraction of microwave signals.  The 'nonlinear port' uses an antenna incorporating a diode (model number 1N4148) as part of a 5 mm by 15 mm rectangular metal loop. Two 'linear' antennas consisting of 5 mm by 15 mm rectangular metal loops are mounted at the other ports (the 'linear port' and 'transceiver port'). The diode is driven by a continuous wave (CW) tone generated (via an HP 83620B swept signal generator) at a frequency fdiode = 400 MHz and a power of +20 dBm; this signal is used to generate intermodulation products with signals incident upon the diode. An alternate realization was also constructed by replacing the nonlinear antenna/diode and the CW tone with a linear antenna connected to a frequency multiplier circuit.  The circuit consists of a Wilkinson divider (HP model 87304C), with both outputs connected to the ports of a x2 frequency multiplier (Mini-Circuits model ZX90-2-50-S+) to form a closed circulating nonlinear circuit which generates second harmonics of the incident signals, without a driving tone. Similar results are obtained with both realizations of the nonlinear port.

\begin{figure}
 \includegraphics{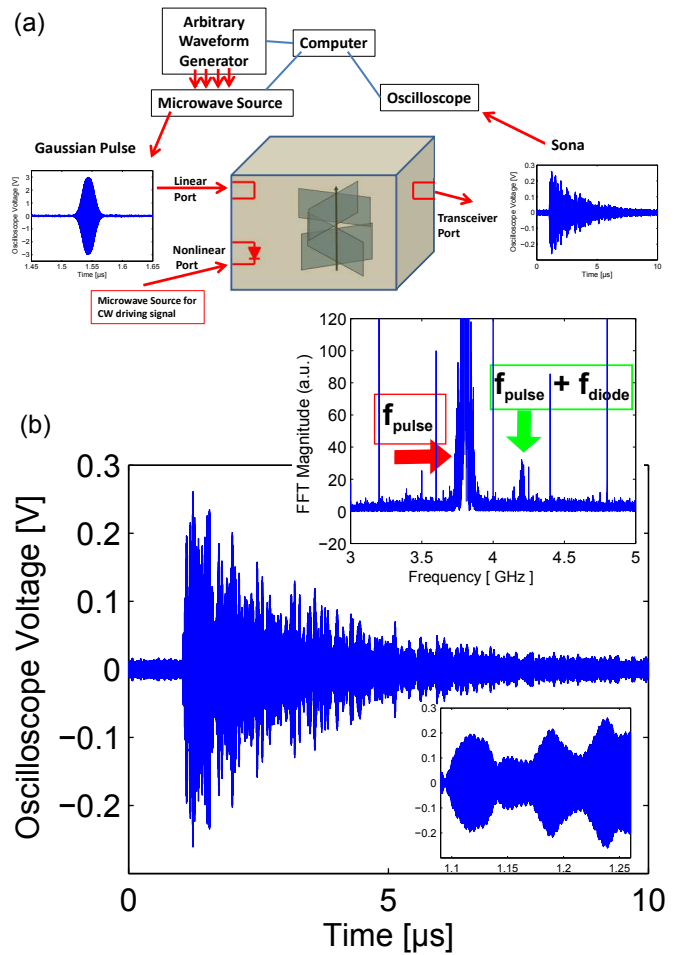}
 \caption{(a) Schematic of the experimental setup. A Gaussian-shaped pulse (fpulse = 3.8 GHz, t = 50 ns) is transmitted into the ray-chaotic enclosure through the linear port, and reverberates through the scattering environment, interacting with the nonlinear element. (b) An example of a full Sona signal measured by the oscilloscope at the transceiver port, including the 3.8 GHz carrier tone and modulation envelope.  (Lower Inset) shows a short segment (150 ns) of the sona in detail. (Upper Inset) Magnitude of the Fast Fourier Transform of the sona shown in (b) as a function of frequency.  The inset shows a close up around the center frequency of the pulse, indicating the frequency components arising from the pulse (fpulse) and the nonlinear element (fpulse + fdiode)}
 \end{figure}

 In the time-forward portion of the experiment, an initial driving signal, consisting of a Gaussian-shaped (in the time domain) pulse, is generated (via a Tektronix AWG7052 arbitrary waveform generator and an Agilent E8267D Vector PSG microwave source) at a carrier frequency fpulse = 3.8 GHz, with a duration of 50 ns and a power of +25 dBm, and is transmitted into the system from the linear port. The excitation propagates throughout the system, including to the diode, where nonlinear excitations are generated as intermodulation products of the pulse frequency and the CW driving signal frequency. The combined signal reverberates through the scattering environment, and is received at the transceiver port and recorded using an oscilloscope (Agilent Infiniium DSO91304A Digital Storage Oscilloscope) over a period of 10 µs, either through a single-shot measurement, or through averaging of several (up to 100) waveforms.
 
  \begin{figure*}
 \includegraphics[width=\textwidth]{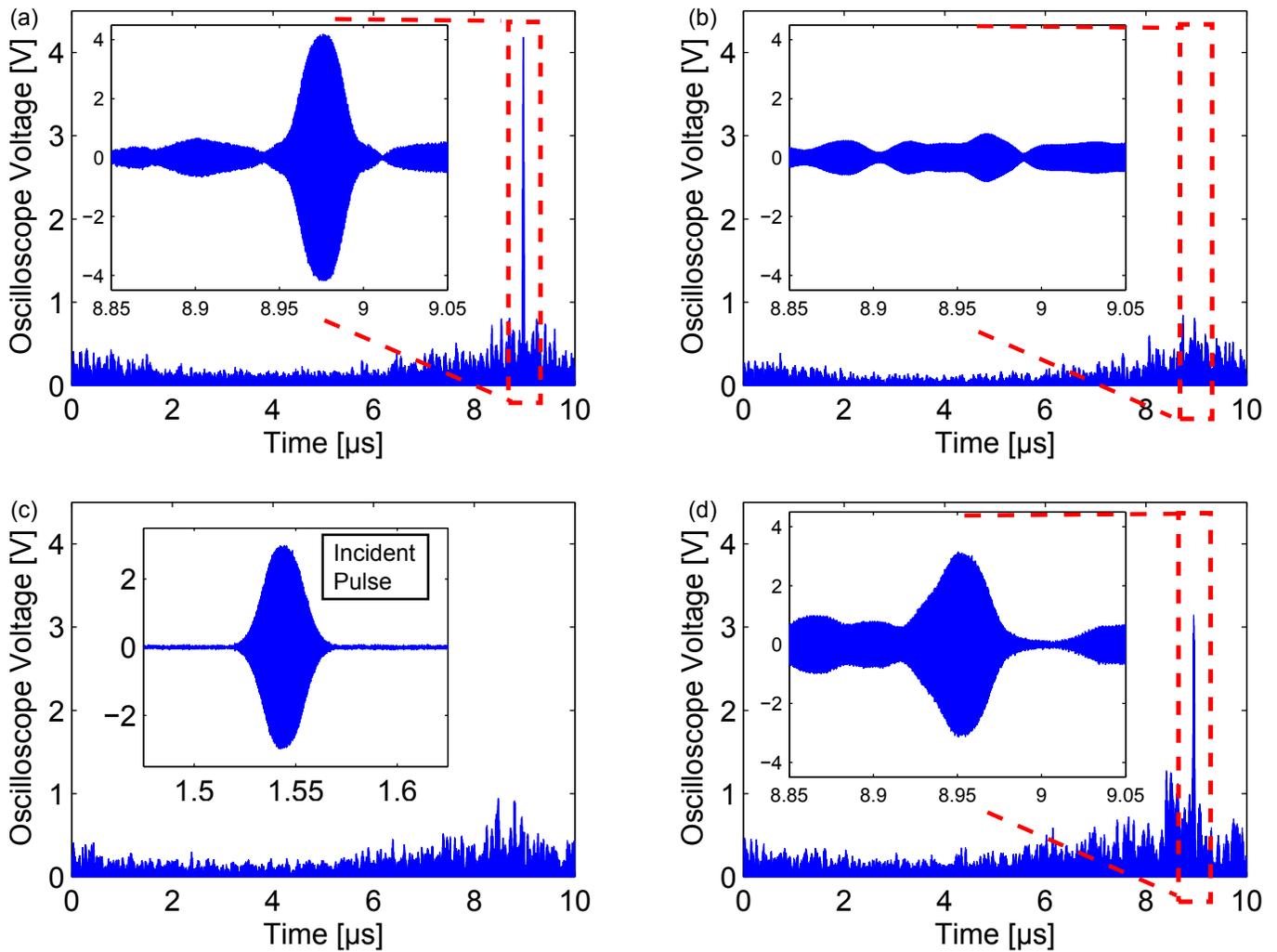}
 \caption{Reconstructions collected at the linear port and at the nonlinear port after broadcast of time-reversed sonas from the transceiver port.  (a) shows the reconstruction of the linear sona at the linear port (only the positive part of the waveform is shown); inset shows the reconstructed pulse in detail.  (b) shows the (lack of) reconstruction of the nonlinear signal at the linear port, while (c) shows a similar (lack of) reconstruction of the linear sona at the nonlinear port. The inset of (c) shows the pulse initially incident on the linear port in the time-forward step. (d) shows the reconstruction of the nonlinear signal at the nonlinear port; inset shows the reconstructed pulse in detail.  }
 \end{figure*}

The recorded waveforms (referred to as 'sonas', an example shown in Fig.  1a and Fig. 1b) are complicated waveforms that are unique to the scattering environment and unique to the source, nonlinear element and detector locations.  An example Fourier transform of such a signal is shown in the inset of figure 1b.  The Fourier transform consists of signals at the carrier frequency, at harmonics of the 400 MHz CW signal, and at the sum and difference frequencies of the two tones, arising from the nonlinear object within the enclosure. (These intermodulation tones are absent when the diode is removed from the antenna. The weak signal at the difference frequency is explained by poor coupling of the antennas at that frequency.) This sona is band-pass filtered into a linear sona (through a filter with bandwidth of 100 MHz centered at the pulse carrier frequency fpulse), and a nonlinear sona at the sum frequency (fpulse + fdiode).

In the time-reversed portion of the experiment, each sona is time-reversed and retransmitted from the transceiver port and the reconstructed signals measured at the linear port and at the nonlinear port. Figure 2 shows example reconstructions measured using this setup. For the linear sona, a reconstructed pulse appears only at the linear port (Fig. 2a) and not at the nonlinear port (Fig. 2c); similarly, for the nonlinear sona, a reconstructed localized-in-time excitation appears only at the nonlinear port (Fig. 2d) and not at the linear port (Fig. 2b). This also demonstrates the exclusive nature of information transfer between the transceiver port and the linear / nonlinear port. 

The question arises about why the nonlinear reconstruction (Fig. 2d) is so well localized in time, in fact resembling the initial pulse. For an initial signal $O$ transmitted into the system at port a, the sona $\sigma$ at port b is expressed as $\sigma_{b} = H_{ab} O$, where $H$ is a transfer function between the two ports. The linear sona generated by excitation at the linear port (LP) at the transceiver port (TP) is expressed as  $\sigma_{linear} = {H}_{LP,TP} O$. The nonlinear sona arises from the signal present at the nonlinear port (NP): $\sigma_{NP}= H_{LP, NP} O$. The nonlinearity generates a new excitation $N = D \sigma_{NP}$, where $D$ is the nonlinear effect on the signal (depending on the properties of the nonlinearity). The received nonlinear sona is then expressed as $\sigma_{nonlinear} = H_{NP,TP} N$. For this realization of the time-reversal mirror, we expect the signal ($\sigma_{NP}$) incident upon the diode (and the resultant nonlinear sona $\sigma_{nonlinear}$) to be dominated by the largest amplitude excitations arising from the direct-path propagation of the initial pulse. This will result in a well-localized in time nonlinear reconstruction. Experiments have been performed to test this hypothesis, and indicate that the sona arriving at the nonlinear port consists of a large amplitude, well-localized initial excitation followed by low amplitude reverberations. When the configuration of the ports is altered to remove the direct path between the linear and nonlinear ports, the initial large-amplitude excitation (and much of the nonlinear sona) disappears. Experiments have also been performed to measure the localization in space of the reconstructions. Using reconstructions upon ports separated by approximately one-half of the pulse wavelength, an upper bound for the nonlinear reconstruction width of $0.96 \lambda$ (of the nonlinear carrier signal) is observed.  Measurement of the size of the received signal on an antenna swept through the reconstructed spatial profile provides a lower bound for the reconstruction width; a full-width half-maximum of $0.77 \lambda$ is observed. These measurements show that the reconstruction is strongly localized in both space and time.

To use the time-reversal of a nonlinear signal as a secure communication channel \cite{a5}, it must not be possible to determine the content of transmitted messages at locations other than the nonlinear port. Using a naïve on-off modulation of the nonlinear sona, it will be possible to decode the signal at any point in the environment, as enhanced noise will be present for 'on' and not present for 'off'.  The communication link can be made clandestine by encoding data as a series of constructed sonas representing '1' and '0' bits. The nonlinear sona received at the transceiver port is utilized to create a '1' bit at the nonlinear object in a pulse code modulation communication scheme.  To create a '0' bit, this nonlinear sona is Fourier-transformed into the frequency domain, and random Gaussian noise is added to the phase information of this signal.  This noisy-phase signal is inverse-Fourier transformed back to the time domain, resulting in a '0' sona that superficially looks like the nonlinear sona to an observer, but does not cause a reconstruction anywhere when it is time-reversed.  The inset of Fig. 3 shows examples of '1' and '0 bit' sonas joined together to form the word '1011'.  A series of these '1' and '0' time-reversed sonas are overlapped by 50\% \cite{a22} (e.g. for a 10 µs sona, an overlap of 5 µs; see Fig. 3 inset) and transmitted at the transceiver port, and the presence or absence of reconstructed pulses are measured at the nonlinear port, and translated into the intended bit pattern. Reception of the nonlinear sona reverberations at other locations in the box will not give information about the bits transmitted to the nonlinear port (as demonstrated in Fig. 2 and Fig. 3).  An equivalent process may be performed with the linear sona, to establish exclusive communication between the transceiver and linear ports.
\begin{figure} 
 \includegraphics{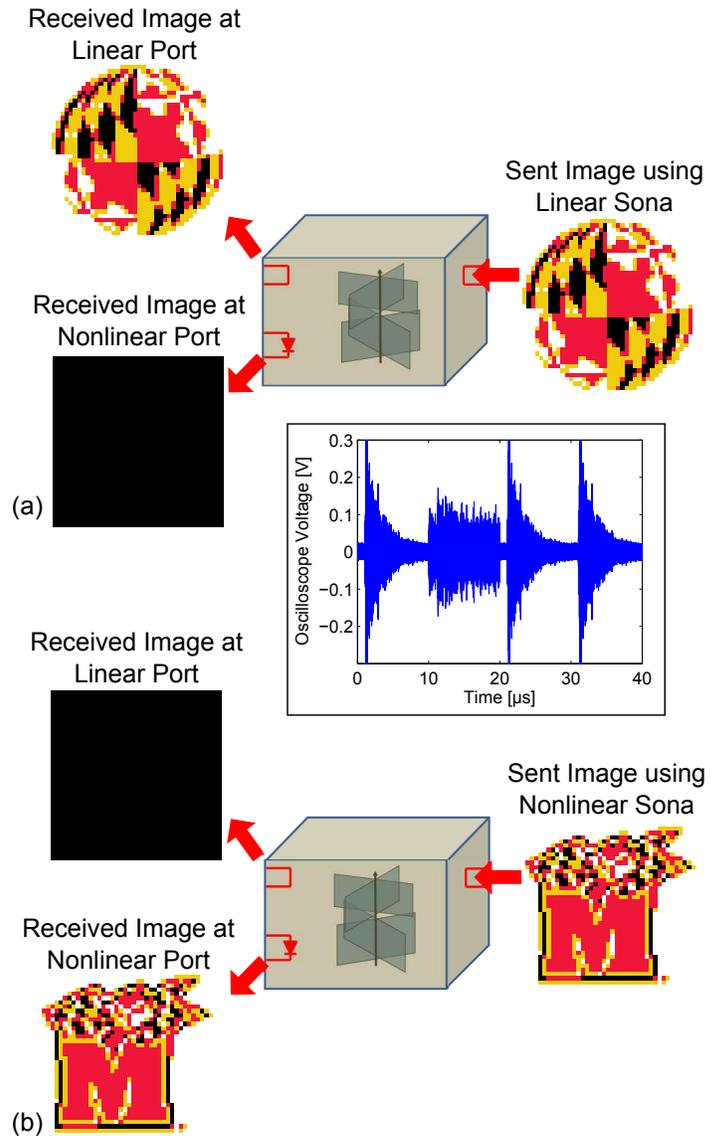}
 \caption{(a) Transmission of a four-color (two bits per pixel) image using the time-reversed linear sona, which is reconstructed only at the linear port. The lack of reconstructed pulses at the nonlinear port corresponds to the transmission of '00' which decodes as a black pixel. (inset) An example constructed sona, displaying the combination of '1' and '0' component sonas, to generate a reconstruction representing '1101'. (b) Transmission of a different four-color image using the time-reversed nonlinear sona, which is reconstructed only at the nonlinear port.   }
 \end{figure}

Figure 3 (a, b) shows images (1600 pixel, four-color) encoded in this manner that were transmitted in an exclusive manner to either the linear port (Fig. 3a) or the nonlinear port (Fig. 3b). The color palette for each image was mapped to four two-bit words (black - '00', red - '01', yellow - '10', white - '11'). For each word, an appropriate sona is constructed from concatenation of '1' and '0' sonas in reverse order (to undo the effect of time-reversal on the message), using the nonlinear sona to address the nonlinear port, and the linear sona to address the linear port.  Prior to image transmission, single linear and nonlinear sonas are transmitted and reconstructed on the respective ports, and recorded as an exemplar reconstruction. The location in time of the exemplar reconstruction is used as a 'clock' to determine a narrow time window in which the reconstructed pulses may appear.  The size of the reconstructed signal determines the detection threshold voltage for the image reconstructions. The constructed sona is transmitted and reconstructs on the appropriate port, and is measured on both linear and nonlinear ports. For each reconstruction time, if the waveform exceeds the threshold voltage a '1' bit is recorded, otherwise, a '0' bit is recorded. The resulting two-bit words are translated back to determine the next pixel color. In Fig. 3a, an image encoded using the linear sona is received with no error at the linear port. At the nonlinear port, the lack of any reconstruction is decoded as '00', appearing as a black image. In Fig. 3b, the converse holds: a different image encoded using the nonlinear sona is decoded without error using reconstructions at the nonlinear port; no reconstructions are measured at the linear port.  Many extensions and improvements of this technique are possible, including the use of linear and nonlinear sonas at the same carrier frequency, greater overlap of the sonas, \cite{a22} and the use of more sophisticated reconstructed waveforms to convey more information.

Our experimental results demonstrate time-reversal of electromagnetic signals arising from a discrete nonlinear element in a wave-chaotic enclosure.  Reconstructions of the linear- and nonlinear- time-reversed signals have been demonstrated to be exclusive to linear- and nonlinear sources, enabling a method for secure communication with the nonlinearity. The ability to 'find' a nonlinear object and exclusively direct signals to it opens up new applications. Using the (possibly amplified) nonlinear sona, high-energy pulses can be reconstructed at a desired location (using a rectenna) forming a wireless power transmission system which avoids using a dangerous high-energy beam for power transmission. Alternatively, the reconstructed pulse could be used for precision hyperthermic treatment of tumors, by applying high-power pulses to nonlinear tags accumulated in the tumor with minimal disruption to other tissue in the scattering environment. Note that in both cases we do not require knowledge of or access to, the location of the nonlinear object, as would be required for linear time-reversal. In addition to the precision of the highly localized pulse (both spatially and temporally), different nonlinear objects may be distinguishable by the spectrum of their nonlinear response, enabling tailoring of nonlinear sonas to focus pulses on specific objects. Furthermore, the time-reversal mirror may be used as a sensor \cite{a11,a12} to detect changes in both the scattering environment (through the linear and nonlinear reconstructions and sonas) and a nonlinear object (through changes appearing only in the nonlinear reconstruction and sona).   The union of time-reversal, wave-chaos, and nonlinear dynamics should continue to stimulate new basic research questions and applications.

Acknowledgments: This work was funded by the Intelligence Community Postdoctoral Research Fellowship Program (20101042106000), the Office of Naval Research AppEl Center Task A2 (N000140911190), the Air Force Office of Scientific Research (FA95500710049), and the Maryland Center for Nanophysics and Advanced Materials.


\begin{thebibliography}{5}
\bibitem{a22} We have resolved distinct reconstructed pulses from pairs of sonas overlapped by 99.5\%, i.e. distinct pulses separated by just one pulse width (50ns in the reported experiments, also demonstrated for 2ns pulses/2ns separations of linear sonas). Overlaps of 4 and 8 sonas have also been demonstrated as suitable for resolving reconstructions.
\end{thebibliography}
\end{document}